\documentclass[floatfix,twocolumn,superscriptaddress,longbibliography,bibnotes]{revtex4-1}
\usepackage{graphics,epsfig,amsfonts,amssymb,amsmath,color}
\usepackage{accents}
\usepackage{bbm}
\usepackage{color}
\usepackage{comment}
\usepackage{enumitem}
\usepackage{natbib}
\usepackage[colorlinks=true,linkcolor=blue,citecolor=blue,urlcolor=blue,hypertexnames=false]{hyperref}
\usepackage{textcomp}
\usepackage{setspace}

\begin{document}

\title{Materials for Silicon Quantum Dots and their Impact on Electron Spin Qubits}

\author{Andre Saraiva}
\email[]{a.saraiva@unsw.edu.au}
\affiliation{School of Electrical Engineering and Telecommunications, \\
The University of New South Wales, Sydney, NSW 2052, Australia}
\author{Wee Han Lim}
\affiliation{School of Electrical Engineering and Telecommunications, \\
The University of New South Wales, Sydney, NSW 2052, Australia}
\author{Chih Hwan Yang}
\affiliation{School of Electrical Engineering and Telecommunications, \\
The University of New South Wales, Sydney, NSW 2052, Australia}
\author{Christopher C. Escott}
\affiliation{School of Electrical Engineering and Telecommunications, \\
The University of New South Wales, Sydney, NSW 2052, Australia}
\author{Arne Laucht}
\affiliation{School of Electrical Engineering and Telecommunications, \\
The University of New South Wales, Sydney, NSW 2052, Australia}
\author{Andrew S. Dzurak}
\affiliation{School of Electrical Engineering and Telecommunications, \\
The University of New South Wales, Sydney, NSW 2052, Australia}


\keywords{Spin qubits, quantum dots, quantum computation, silicon nanostructures, quantum devices}

\begin{abstract}

Quantum computers have the potential to efficiently solve problems in logistics, drug and material design, finance, and cybersecurity. However, millions of qubits will be necessary for correcting inevitable errors in quantum operations. In this scenario, electron spins in gate-defined silicon quantum dots are strong contenders for encoding qubits, leveraging the microelectronics industry know-how for fabricating densely populated chips with nanoscale electrodes. The sophisticated material combinations used in commercially manufactured transistors, however, will have a very different impact on the fragile qubits. We review here some key properties of the materials that have a direct impact on qubit performance and variability. 
\end{abstract}

\maketitle

\section{Introduction}

Reaping the benefits of quantum computing for the solution of real-world problems will require integrating a large number of high quality qubits. Silicon stands out as a unique material in this sense. Besides being the standard material for very large-scale integrated microelectronics, research shows that the spins of its electrons have excellent qubit properties, such as long coherence time and efficient individualised controllability. Much of what is known about quantum computers, however, is still the product of academic research -- only recently industrial players started showing significant progress towards building quantum processors. For semiconductor-based quantum technologies, this shift to foundry-fabricated devices represents a major cornerstone, vastly expanding the repertoire of processes and three dimensional integration made possible by the advanced engineering techniques developed for classical transistors. 

Understanding the role played by various materials in a device stack will be paramount. University fabrication facilities typically have a narrower range of options in terms of device processing, but reasonable flexibility in adapting process steps between device generations. On the other end of the spectrum, foundries have well-established sophisticated processes that are highly optimised and reproducible, but changes in the process flow can be difficult.

In this context, we review the scientific progress in fabricating qubits based on electron spins in gate-defined silicon quantum dots~\cite{loss1998quantum,zwanenburg2013silicon,ladd2018silicon,zhang2018qubits,laucht2021roadmap}. These devices have the potential to leverage the most advanced form of semiconductor engineering from the complementary metal-oxide-semiconductor (CMOS) industry. We focus on the choices of materials and what is known about their impact on qubit performance, substantiated by decades of academic research. These performance benchmarks often reveal that the key material properties of relevance for spin-based quantum computing are different from those for classical transistor devices, and adapting to these differences will be amongst the most important challenges for the quantum ambitions of semiconductor manufacturers in the upcoming years.

We note that recent progress has also been achieved using hole-based qubits~\cite{camenzind2021spin, maurand2016cmos,hendrickx2020singlehole,hendrickx2021fourqubit}. Holes have certain advantageous properties, including the easily lifted degeneracy of the valence band (as opposed to the valley degeneracy of the conduction band, discussed later) and the possibility of performing electrically driven spin resonance (EDSR) leveraging their naturally large spin-orbit coupling. The impact of materials on hole qubits is significantly different from electron qubits -- even the angular momentum of the resulting  hole-based two-level system can change from $\pm 1/2$ to $\pm 3/2$ depending on strain and confinement geometry~\cite{terrazos2021theory}. We therefore will not discuss materials impacts on hole-based CMOS qubits here, but will leave this as the topic of a future review.

This review is organised in six sections. Sections 1 and 2 give an introduction of the current status of quantum computing and how spins in silicon quantum dots can be used as qubits. We conclude the introductory two sections of the review with some comments on the scalability of this system and the current challenges. The next sections are a revision of the impact of materials on qubit devices. Section 3 gives an overview of which choices of materials have been explored so far and their consequences on qubit quality. Section 4 discusses the sources of electromagnetic noise in these materials and what is known about their impact on qubits. The variability between qubit parameters is the theme of Section 5. In Section 6 we draw some conclusions based on the overview presented here, and present some projections for the future of this technology.

Our review is particularly focused on the aspects of relevance in deciding the choices of materials in an integrated device fabrication process and how those would impact the quantum processor. Other reviews on the general topic of silicon-based quantum computing have focused on different aspects of the technology. Examples include Ref.~\cite{zwanenburg2013silicon}, which discussed the physics of various nanostructures that can be used to create spin qubits, Ref.~\cite{vandersypen2017interfacing} focused on the full stack view of the quantum processor, and Ref.~\cite{leon2021materials} reviews the issue of materials for general quantum technologies but does not go into the details of the complex combination of materials in modern foundry processes.

\subsection{Current Status of Quantum Computing}

Quantum processors for gate-based computations are now routinely fabricated, with sizes ranging from one to tens of qubits \cite{arute2019quantum,egan2021faulttolerant,saffman2016quantum,xu2020demonstration,hendrickx2021fourqubit}. These proof-of-principle devices provide a test-bed for some of the theoretical questions that will need to be answered to realise a full scale quantum computer. The first obvious conclusion drawn from these early tests is that the errors introduced by noise at the quantum level build up quickly~\cite{arute2019quantum}, as foreseen theoretically decades ago~\cite{preskill1998reliable}. While these faulty quantum processors may find immediate purpose-specific applications, in what is dubbed noisy intermediate-scale quantum (NISQ) computing~\cite{preskill2018quantum}, universal quantum computation will not be achievable without correcting for these errors.

Quantum computers will need to evade these errors with a scheme to encode logical qubits redundantly in a collection of physical qubits. The rules of quantum mechanics, however, set stringent overheads for efficient error correction. Current quantum computers are very far from the scale needed to perform universal quantum computation that is fully error corrected -- at best, such a task would require tens of thousands of high-performance qubits~\cite{barends2014superconducting}. This number becomes even larger once the physical limitations in qubit connectivity in a two-dimensional array are taken into account.

Early, small-scale demonstrations of the key elements for quantum error correction are promising. Using ion-trap processors, experimental demonstrations of a 9-qubit error correction code recently achieved fault tolerance levels (in which the collective behaviour of the logical qubit outperforms the individual physical qubits)~\cite{egan2021faulttolerant}. Another promising early demonstration of a two-logical-qubit operation using a technique called lattice surgery~\cite{erhard2021entangling} clarifies the path for a complete gate-set for logical qubits. However, in order to achieve useful large scale quantum computations, the techniques used in these experiments would have to be adapted for scalability. 

\subsection{General Challenges for Scaling Up Quantum Processors}

Classical processors typically consist of billions of transistors~\cite{IRDS}. In order to achieve this, challenges in transistor miniaturisation, heat dissipation and wiring had to be overcome. These challenges, when translated to the quantum realm, become significantly more striking.

With relation to sizing, the array of quantum processing elements needs to be monolithically integrated on a chip in order to achieve the density necessary for operating millions of qubits simultaneously. This limits the maximum area of the quantum processor to roughly the size of the milikelvin stage of a dilution refrigerator, which typically lies in the tens of centimeters. In order to cover this area with the number of qubits and the auxiliary electronics required for qubit operation and measurement, miniaturisation becomes a key aspect.

Heating, which is a key limitation for classical bits, is a significantly worse problem for quantum computation since most qubits require sub-kelvin temperatures to operate. In solid-state devices, cooling the quantum processor and associated control chips with a dilution refrigerator imposes a bottleneck since electronic devices generate heat at a rate that surpasses the refrigerator cooling power. This issue may also impact qubits leveraging microfabricated ion traps on a chip. The electric field generated by the electrodes communicate thermal vibrations from the chip to the ions~\cite{monroe2013scaling}, limiting qubit coherence and potentially impacting the maximum number of ions that can be held in a single trap.

Perhaps the most daunting difficulty relates to the wiring required for individualised qubit control. Quantum computer operations typically involve connecting each qubit to some classical electronic device, such as an arbitrary waveform generator. This creates difficulties related to the bonding of these wires to a chip~\cite{franke2019rent}, the physical dimensions of the dilution refrigerator bore, and the sheer amount of classical electronic devices needed for performing complex computations. Moreover, coupling separate large arrays of qubits might require the development of quantum links between separate modules, in order to reduce the density of qubits and to allow for the interspersing of auxiliary electronic devices~\cite{vandersypen2017interfacing}. Classical computers rely on multiplexing and integrated circuits to control the billions of classical bits, and the pathway to adapt these technologies for quantum control are not yet established.

Note that quantum links, or more generally a quantum network, will rely on the coherent transfer of quantum information and is, in itself, a very challenging task~\cite{gold2021entanglement,daiss2021quantumlogic,zhong2021deterministic}. These links are important not only to allow room for incorporating control electronics, but also for improving the connectivity of the qubit arrays. They may also be used to connect qubits in separate chips (instead of monolithically integrated qubits), but this only represents a minimal gain in terms of scalability. This gain may be important for small scale demonstrations of NISQ computations, but does not solve the problem of the overhead required for quantum error correction.

In order to construct a fault-tolerant quantum processor with the device density needed to realize millions of qubits in a compact space, spin qubits based on quantum dots that can leverage CMOS integrated circuit manufacturing provide a promising way forward. In the next section we review the key conceptual and operational requirements for such qubits.

\section{Electron Spin Qubits in Silicon Quantum Dots}

Spins of electrons are natural two-level systems, such that encoding $\left| 0\right\rangle$ and $\left|1\right\rangle$ is straightforwardly done by considering the $\left|\uparrow\right\rangle$ and $\left|\downarrow\right\rangle$ states, respectively~\cite{loss1998quantum,kane1998siliconbased,nielsen2010quantum}. This encoding of quantum information in spin states is common across many platforms including atoms in traps and defects in diamond. We focus here on the case where the spins are isolated, manipulated and measured in electrostatically defined quantum dots~\cite{loss1998quantum} based on silicon. Reference~\cite{stano2021review} provides an excellent comparison of the performance metrics of spin qubits in gated semiconducting nanostructures in general.

An electrostatic quantum dot typically consists of a stack of materials~\cite{zwanenburg2013silicon,hanson2007spins} that is designed to create a quantum dot by combining electric fields and interfaces between materials. The electron is confined at the interface between the active material -- which we specify as being silicon for this paper -- and some barrier material with a conduction band offset compared to silicon. The most common choices of barrier materials are SiO$_2$ (forming a standard MOS device) or an alloy of silicon and germanium (Si$_{1-x}$Ge$_{x}$).

For these spins to be considered potential qubits, it is necessary to be able to isolate a single spin, control it, entangle pairs of spins and finally read them out. Moreover, all these tasks must be performed with high precision and fast enough to avoid the impact of electromagnetic noise. These criteria were summarised by David DiVincenzo for a general qubit system, and are considered the bare minimum for a two-level system to be considered a potential qubit~\cite{divincenzo2000physical}.

\subsection{Isolating a Two-Level System}

While spins of electrons are naturally a two-level system, it is important that no other degree of freedom couples to the spin of the electron. This is achieved by providing strong confinement, guaranteeing that the orbital and valley energy levels are well spaced. For a single electron in silicon, the critical excitation energy is set by the valley structure of the conduction band. 

Silicon has a non-trivial conduction band, with six different minima that correspond to Bloch states having the same energy (referred to as valleys). These states consist of waves travelling along the $\pm x, \pm y$ and $\pm z$ cartesian directions, set by the cubic symmetry of the crystal. Their crystalline wavenumber is $k_0\approx 0.85\times 2\pi/a_0$, where $a_0=0.543$\,nm is the lattice parameter of silicon.
In a quantum dot device, this degeneracy is lifted by various effects, including the strain fields, asymmetry of the confinement potential and the geometry of the silicon/insulator interface. However, the energy splitting between the Bloch states associated to electrons with opposite wavevector is in general much smaller than any other excitation energy. It is important, therefore, to understand in depth the physical origin of this energy separation and how to optimise it.

Electrons with opposite wavevectors are coupled by electrostatic potentials that vary abruptly on the scale of the wavelength~\cite{saraiva2009physical}. Since this wavelength is only $a_0/0.85\approx0.64$\,nm, this means that an atomically abrupt scattering potential is necessary. This is typically provided by the interface between silicon and insulator, which then needs to be almost atomically flat. The resulting energy splitting is less than a millielectron-volt, emphasizing the prominent role that valleys play in the low energy physics of electrons in quantum dots. 

Realistically, no material interface is perfectly flat. In order to have a purely spin-based two level system, it is then necessary to engineer the interface to minimise roughness. Typically, the two spin states are separated in energy by the Zeeman splitting $\Delta E_\mathrm{Z}=g_\mathrm{eff}\mu_\mathrm{B} B_0$, where $B_0$ is an external DC magnetic field, $\mu_\mathrm{B}$ is the Bohr magneton and $g_\mathrm{eff}$ is the effective Land\'e factor as affected by the material's spin-orbit coupling. This sets the energy scale that needs to be preserved, determining the minimum workable valley excitation energy. Throughout this review we describe qubits in the regime where the valley excitation is well separated from the spin degree of freedom. Other regimes of strong spin-valley mixing may be of interest, for example, for fast initialisation~\cite{yang2013spinvalley} and spin-valley electrically-driven resonance~\cite{hao2014electron,corna2018electrically}.

While in this review we focus on a single spin as a qubit, other qubit encodings are possible by leveraging the spin degree of freedom in multielectron states ~\cite{levy2002universal,laird2010coherent,kim2014quantum}. Each of these encoding schemes has their own advantages and disadvantages and is impacted by the valley degree of freedom in different ways.

\subsection{Spin Readout}

Qubit readout is one of the most demanding tasks for a quantum computer, often overlooked when comparing qubit technologies. It is important to be able to measure qubits in a fast, accurate and scalable way. Quantum error correction requires single-shot readout of qubits at fidelity levels comparable to all other qubit operations, \textit{i.e.}, with error rates well below one in a hundred shots. Moreover, qubit measurements need to be performed several times before any error occurs, which means that the readout bandwidth must be much faster than the qubit decoherence times. Finally, readout of one qubit must not deteriorate the operation of the neighbouring qubits.

The magnetic dipole of electrons is far too small for detecting the spin directly. Instead, spin-dependent charge movement is necessary in order to distinguish qubit states. In this way, instead of measuring the spin, one measures the electron position, which reveals what the original spin state was. 

From a microscopic point of view, there are several different ways by which this type of spin-dependent transition may occur, but they usually rely on either the Pauli spin blockade~\cite{ono2002current, fogarty2018integrated,seedhouse2021pauli} or the Zeeman energy splitting between two spins states under an externally applied magnetic field $B_0$~\cite{elzerman2004singleshot,veldhorst2014addressable}. While the latter requires that the magnetic field is sufficiently large to create a significant splitting between spin states (typically larger than the thermal energy), the former only requires enough magnetic field to prevent spin-orbit induced singlet-triplet mixing.

The charge movement may then be detected using a range of methods, the simplest one being to detect transitions by measuring current passing directly through the quantum dots that harbour qubits~\cite{ono2002current}. This technique can only measure the average result of many repetitions of the experiment, since the passage of a large number of electrons is required in order to obtain a measurable current. Nevertheless, this has been used in order to demonstrate coherent control of spin qubits in simplified quantum devices~\cite{maurand2016cmos,camenzind2021spin}.

One way single shot readout can be obtained is by measuring the current going through a different dot, which is remotely impacted by the charge state of the qubit quantum dot~\cite{knobel2003nanometrescale,morello2010singleshot}. This remote quantum dot is usually referred to as single electron transistor (SET), and can be tuned to be maximally sensitive to shifts in electrostatic environment. This is achieved by setting a very small difference between the chemical potentials in the source and drain of the SET and aligning one of the charge transitions to lie in the window between these two energy levels. Small changes in electrostatic environment will corrupt this alignment, leading to a measurable change in current~\cite{reilly2007fast}.

In its simplest form, the DC current is sufficient to distinguish remote charge states from random electric noise (in devices with less electric noise, such as GaAs, even a simpler quantum point contact is sufficient for detecting these charge movements~\cite{field1993measurements}). The signal-to-noise ratio and measurement bandwidth can be significantly improved, however, by measuring the response of the SET to radio-frequency excitations (RF-SET)~\cite{schoelkopf1998radiofrequency,angus2008silicon}. This is quite important in the context of quantum error detection, since a better signal-to-noise ratio will allow for faster identification of potential errors in the computation.

The use of an on-chip SET consumes a large area near the active region where quantum dots are defined. Moreover, the electrostatic interaction with the quantum dots decays with distance, such that for a full scale quantum computer it would be necessary to fabricate one of these SETs for every few qubits. These difficulties have led to the development of gate-based readout methods~\cite{colless2013dispersive,betz2015dispersively,west2019gatebased}. In this case, a microwave excitation is applied directly to one of the gate electrodes in the device and the reflected amplitude is measured. 

If the charge arrangement within the quantum dots is stable, the small amplitude of the microwave excitation creates no charge movement. Otherwise, if the voltage biases are arranged such that any of the electrons is on the brink of transitioning between different dots (or from a dot to a reservoir), the microwave excitation will push the electron back and forth, such that the quantum dot acts as a small capacitor. Small changes in capacitance can be measured by engineering a resonator with very high $Q$ factor, such that small shifts in its resonant frequency can be detected. This technique is referred to as gate-based dispersive readout~\cite{delbecq2011coupling}. We return to the topic of resonators with large $Q$ factors in Section 3.

Reflectometry techniques have the advantage that no additional resources are required at the qubit level -- the same gate that was used to form and control the quantum dots is now used to measure the spins. 

Overall, single-shot spin readout fidelities of $>90$\% are routinely achievable w ith different types of readout methods~\cite{stano2021review}. However, to achieve fault-tolerant readout with $99.9$\% fidelity, quantum non-demolition readout has to be employed, where the state of the qubits gets repeatedly transferred to and read out via a neighbouring qubit~\cite{xue2020repetitive,yoneda2020quantum}.

\subsection{Spin Initialisation}

Quantum computations start from a known initial state of the qubits, which requires the development of a strategy to initialise all the qubits. In its simplest form, flushing the electron out of the quantum dot and loading another electron into the quantum dot ground state (separated from the spin excited state by the Zeeman splitting) is an effective strategy for initialisation~\cite{elzerman2004singleshot}, but requires the dot to be near a reservoir. For a dense arrangement of quantum dots, only the dots at the perimeter of the array can be connected to a reservoir (unless 3D integration of reservoirs is made possible through vias~\cite{cai2019silicon}). It is, therefore, desirable to create a route for initialisation that does not rely on coupling most qubits to a bath of electrons.

Initialisation may in general be achieved based on relaxation. The natural route for relaxation of an isolated spin in silicon is the inelastic scattering of phonons, creating an oscillatory electric field which couples to spins through the spin-orbit interaction~\cite{tahan2005rashba}. The rather indirect coupling between spins and phonons and the fact that this effect requires a pre-existing phonon conspire to lend the spin a long relaxation time, of the order of hundreds of milliseconds in silicon. Normally considered a desirable property for qubits, this is a significant limitation in terms of initialisation time and fidelity. 

This process can be engineered to be faster by controlling the orbital or valley energy splitting to match the Zeeman energy, creating what is known as a relaxation hotspot. The limits for hotspot initialisation are unclear, but voltage-tunable relaxation as fast as microseconds has been achieved~\cite{yang2013spinvalley}.

An isolated pair of spins in a double dot can also be initialised by moving both spins to the same site and relying on the relaxation to a singlet state. This initialisation involves a spin flip process as well, but it is in general faster than the single spin relaxation because the energy separation between triplets and singlets on the same dot is larger than the Zeeman splitting. Then, by ramping the potential detuning between the two dots back to the configuration with separated spins, either a singlet or an anti-parallel spin configuration can be obtained. What determines the final state is how the ramp rate compares to the splitting induced by the difference in Zeeman energies in the two dots, which determines the level of adiabaticity of the process.

Another form of initialisation is by quantum non-demolition measurements~\cite{nakajima2019quantum,yoneda2020quantum,xue2020repetitive}. In general, measuring a qubit generates a random back-action on its state. There are, however, strategies by which the target spin qubit can be entangled to another spin, and this ancilla spin is then read out. Engineering the interaction between the ancilla and the target qubit to be spin conserving, all the random measurement back-action is steered away from the spin projection, leading to a collapse of the wavefunction into the state corresponding to the measurement outcome. The collapsed new state is completely determined, so the qubit can now be considered to be initialised. If initialisation in a particular spin state is required, one can manipulate the qubit spin conditional on the outcome of its measurement~\cite{nakajima2019quantum}.

This technique is particularly appealing as a method for high fidelity spin readout because it allows for repetitive measurement, which can be used for real time Bayesian estimation of the spin state~\cite{yoneda2020quantum}. Note that the strategies to measure the ancilla qubit will be the same as those discussed in the previous session, imposing similar requirements in terms of measurement apparatus.

Interestingly, spin initialization is probably the least well quantified performance metric of spin qubits and usually resides in the few percent range~\cite{stano2021review}. However, initialisation by measurement should in principle be able to make the initialization fidelity equal to the measurement fidelity.

\subsection{Universal Gate Set}

In order to perform universal quantum computation, it is sufficient to be able to implement arbitrary single qubit rotations and two qubit entangling gates~\cite{divincenzo2000universal}. These gate operations need to be highly controllable, fast compared to the decoherence rate and capable of addressing a target qubit without disturbing other qubits in an array.

Spins in silicon were recognised as potential high-performance qubits at an early stage~\cite{kane1998siliconbased}. Silicon has a relatively small spin-orbit coupling, no piezoelectric phonons and naturally abundant spin-zero isotopes, which makes for a relatively undisturbed magnetic environment for the spins. These can be further improved by isotopic purification~\cite{abe2010electron} and controlling the spin-orbit coupling~\cite{tanttu2019controlling}.

Control of a single spin in a magnetic field is in general obtained through spin resonance methods. Applying oscillatory electromagnetic fields, it is possible to create spin rotations whenever the frequency of the driving field is near the natural precession frequency of the spin qubit (also called the Larmor frequency). The amplitude of the oscillatory field determines how fast the spin rotates in a nutating movement. The frequency at which the spin nutates when driven resonantly is called the Rabi frequency.

It is typical to target either predominantly magnetic or electric fields when designing the source of the oscillatory resonant field. A purely magnetic field directly couples to the spins, while an electric field will require an indirect coupling. In a typical electrically-driven spin resonance, a micromagnet is positioned near the dot system, creating an inhomogeneous magnetic field. By physically moving the quantum dot potential well back and forth -- which can be achieved by applying an oscillating voltage bias to one of the gate electrodes -- the electron experiences an effective time dependent magnetic field~\cite{pioro-ladriere2008electrically}. Magnetically-driven spin resonance is typically achieved by creating a time-dependent magnetic field by passing an oscillating current through a metallic nanowire fabricated in the vicinity of the dots~\cite{koppens2006driven}. Other possible methods for driving single spins include the use of dielectric resonators~\cite{vahapoglu2021singleelectron} and electrically driven spin resonance using the strong spin-orbit coupling near the spin-valley degeneracy~\cite{corna2018electrically}.

The desirable qubit properties that spins in silicon have are reflected in the high fidelity operation of qubits~\cite{stano2021review}. Single qubit operations with fidelities exceeding 99.9\,\% have been demonstrated by both electrically~\cite{yoneda2018quantumdot} and magnetically~\cite{yang2019silicon} driven spin resonance. 

Coupling of two spin qubits for performing entangling gates can be achieved leveraging the natural exchange coupling between electron spins whenever their wavefunctions overlap with each other~\cite{divincenzo2000universal}. This overlap can be controlled by tuning the potential barrier between adjacent quantum dots.

Two-qubit gates leveraging exchange coupling have been implemented experimentally~\cite{xue2019benchmarking,zajac2018resonantly,huang2019fidelity}, with fidelities measured by randomized benchmarking up to 98\,\%~\cite{huang2019fidelity}. This fidelity will need to be improved further, preferably well above 99\%, in order to enable fault-tolerant quantum computing. Strategies to improve two-qubit fidelities include increasing the speed of exchange gates, pulse engineering, and the optimization of materials, which is the focus of the latter sections of this review.

\subsection{Flying Qubits}

We note that exchange coupling is a contact interaction. Implementation of logic gates between qubits that are a long distance apart would require swapping the quantum state of neighbouring spins by performing exchange-based operations, such that the quantum information associated with a particular qubit can be shuffled around in a spin array~\cite{kandel2021adiabatic}. The problem is that each operation carries with it a risk of errors. Also, two-qubit gates require individualised quantum dot coherent control, which creates a demand for densely controllable quantum dot arrays. For this reason, it is valuable to explore alternative methods by which quantum information can be transported between different regions of the processor, via what are generally termed flying qubits.

The most evident approach is to simply move electrons around. This can be done by controllably tunnelling electrons between empty dots~\cite{fujita2017coherent,buonacorsi2019network}, for which experiments in silicon quantum dots have shown information transfer fidelity averaging 99.4\,\%~\cite{yoneda2020coherent}. The electrons could in principle be pumped using electrostatic pulses across a series of dots in a bucket brigade manner, without the necessity for individualised quantum gates. Another strategy to move spins around is to leverage the interaction between electrons and surface acoustic waves. This latter strategy has not been tested in silicon devices, but early successes in GaAs-based electrostatic quantum dots are encouraging~\cite{hermelin2011electrons}.

Another direct extension of the technology employed for electrostatic quantum dots is to use a mediator dot. The spin structure of the mediator dot is communicated between remote dots, creating an effective long range interaction between qubits~\cite{fedele2021simultaneous}. This creates some limitations on the design of the qubit array, consuming chip real-estate to form the large mediator dots, but it conversely allows the qubit-bearing quantum dots to be spaced out, potentially improving their connectivity with external wiring.

A potentially more significant shift in paradigm could be offered by coupling spins to microwave photons in a superconducting resonating cavity~\cite{mi2018coherent,samkharadze2018strong}. Photons can travel long distances and allow for a true quantum network, potentially connecting qubits in different regions of a chip, different chips and even different cryogenic setups~\cite{magnard2020microwave}.

\subsection{Advantages of Spins in Silicon for Scaling Up}

The clearest advantage of silicon-based qubits from the perspective of scalability is the prospect of leveraging microelectronics industry know-how for densely populated integrated circuits. This plays a role in integrating the quantum layer of qubit devices with the necessary classical processing systems~\cite{veldhorst2017silicon,pillarisetty2018qubit} responsible for measuring ancilla qubits to check the sanity of the computation, identifying potential errors and acting accordingly to correct those errors. Moreover, circuit integration is important for multiplexing the control of qubits~\cite{li2018crossbar} and alleviating the unmanageable growth of the number of control wires with the number of qubits~\cite{franke2019rent}. 

Other advantages can only be appreciated from a more technical perspective. One example relates to the dominant class of quantum errors present in spin qubits. A single spin has much longer relaxation times than coherence times. This means that there is a large bias towards phase-flip errors, whereas bit-flip errors are typically a thousand times less frequent. This creates an opportunity for bespoke quantum error correction codes that can cope with much higher physical error rates than in the usual case of unbiased errors~\cite{bonillaataides2021xzzx}.

Quantum processors with upwards of a million qubits, as required for quantum error correction, will require considerable readout and control electronics, either on-chip, or closely integrated, which will dissipate considerable power. It would therefore be highly advantageous to operate at higher temperatures, preferably above one kelvin. At these temperatures, cooling can be performed far more efficiently through adiabatic cooling via pumped $^4$He, in contrast with the very limited cooling power of the dilution of $^3$He in $^4$He required for sub-kelvin refrigeration. Spins in silicon have been demonstrated to operate well at these higher temperatures~\cite{yang2020operation,petit2020universal}.

Still on the topic of improved error correction strategies, the most efficient encoding schemes use the pairwise interaction of distant qubits~\cite{richardson2003error}. This interaction at a distance is normally difficult to implement. Electron spins have the advantage that they can be coherently transported with high fidelity~\cite{yoneda2020coherent}. This creates the potential for a bucket-brigade scheme of moving qubits around, as noted earlier.

\subsection{Current Challenges for Silicon Qubit Scalability}

Even though silicon qubits themselves are small and can be easily scaled up by mere replication, current spin qubit technology relies on ancillary devices for initialisation, control and readout of the qubits with much larger footprints than the quantum dots themselves.

The minute size of spin qubits is a double edged sword. In order to confine individual electrons and control their interaction, quantum dot devices in silicon require the fabrication of small gate electrodes with technologies employed in the most aggressively miniaturised transistor devices. Progress in qubit demonstrations has therefore been limited by challenges in nanofabrication and device yield, especially in the context of academic fabrication facilities. 

Initialisation and readout based on spin-dependent tunnelling to a reservoir with a large number of electrons is problematic because these reservoirs typically need to be connected to an external voltage source or ground in order to remain at a fixed chemical potential. This challenge can be alleviated by using Pauli spin blockade for readout and hotspot initialization, completely removing the need for reservoirs during operation. We refer to this reservoir-independent quantum dot operation as an \emph{isolated mode} of operation~\cite{bayer2019charge,yang2020operation}.

There are also open questions regarding the stability of operating a large array of dots in isolated mode, without the presence of a nearby reservoir to stabilize the chemical potential. This could lead to random charging/discharging events in the dot system, potentially requiring reinitialisation of the complete dot array.

Moreover, readout based on spin-dependent tunnelling (regardless of whether it is to a reservoir or between quantum dots) has to be followed by a charge measurement. This is routinely performed using either an (on-chip) SET for charge sensing or a (typically off-chip) resonator for reflectometry. Both techniques require extra chip real estate. Potentially with some form of multiplexing, these resources could be shared among groups of physical qubits~\cite{veldhorst2017silicon}. Moreover, quantum error correction schemes that are less based on real time readout of syndromes~\cite{ercan2018measurementfree} could be an important development to relax the need for high fidelity individualised qubit readout.

Finally, spin resonance is typically performed in one of two ways -- through an oscillating magnetic field driven by a microwave stripline antenna or by electrically driving the dot to oscillate under the inhomogeneous magnetic field of a small on-chip magnet. Both techniques require the magnetic field (AC for the antenna, DC for the magnet) to be generated at a small distance from the quantum dots. Meanwhile, both an antenna or a magnet are much larger than the size of the dots themselves. So far this problem has been circumvented by using a shared antenna or magnet to drive simultaneously a few qubits, addressing them separately by their difference in resonance frequencies. This strategy will encounter hurdles as the number of qubits increases, generating qubit crosstalk, frequency crowding and the inability to simultaneously drive multiple spin qubits. For magnetically driven spin resonance the solution is likely to involve a globally applied field~\cite{vahapoglu2021singleelectron}, whereas for electrically driven spin resonance it will be necessary to employ nanoscale magnet arrays, perhaps as currently employed in magnetic memories.

\section{Overview of Material Choices}

The choice of materials for fabricating quantum processors based on electrostatic quantum dots has played a major role in the success of spin qubits in silicon, and the physics behind the interplay between these materials and the behaviour of single electrons is now better understood. With the increasing participation of industry players, significant knowledge from CMOS chip foundries will now translate to qubit production, opening new exciting prospects.

\subsection{Materials Issues going from Classical to Quantum Bits}

Despite the advanced knowledge about impact of materials on classical transistors, qubits impose entirely new constraints on materials parameters. These are related to three major differences between the operation of classical and quantum devices:

\begin{enumerate}
    \item The temperature of operation for quantum dot devices is 4\,K or lower, while most classical transistor technologies are designed for room temperature operation;
    \item Only a single or a few electrons are isolated in a quantum dot, so that screening of charge noise and electrostatic disorder is far less efficient than in classical transistors;
    \item Spin is susceptible to different types of noise and disorder compared to electric current, which means that the material stack for classical transistors needs to be reassessed for spin qubits.
  
\end{enumerate}

We now elaborate on these challenges and why even the most advanced semiconductor fabrication platforms will require further process development in order to fabricate qubit-grade quantum dots.

\subsubsection{The Transition from Room Temperature to Nearly Absolute Zero}

The challenges of operating semiconductor devices at the required low temperatures for quantum computing start with reaching and maintaining such temperatures during operation. Typically, a dilution refrigerator is needed, which is an expensive piece of equipment and creates significant limitations in terms of space and access to the device. The worldwide availability of $^3$He is also quite limited~\cite{shea2010helium3}. But the most important limitation relates to the cooling power of these refrigerators -- the dilution principle is useful in reaching sub-kelvin temperatures but is an inefficient method of cooling, which limits the maximum power that the device can dissipate to roughly 1 mW.

The next difficulty is the performance of semiconductor devices at low temperatures. Most commercial devices are designed for operation at less extreme temperature conditions, so specialised amplifiers, CMOS controllers and other devices need to be purpose-built~\cite{hornibrook2015cryogenic,patra2018cryocmos}. Incomplete ionisation of dopants below around 50K reduces the flexibility of doping strategies. Finally, electrons in one or two-dimensional arrangements (such as nanowires or planar MOSFETs) will become localised even with minimal disorder due to Anderson localisation~\cite{ando1982electronic}.

These low-temperature localisation effects due to disorder also result in more stringent conditions on the gate sizes. For an electron to be bound laterally by a controllable electrostatic potential, the gate electrode must be made smaller than the randomly located puddles of charge that form in a disordered potential profile.

Another challenge created by the extreme temperature is the strain created by differences in thermal expansion coefficients. Firstly, there is limited knowledge about the temperature dependence of this coefficient and how it behaves at low temperatures. Secondly, the mismatch between materials can cause very inhomogeneous strain fields, impacting the design of quantum dot structures and potentially forming strain-induced dots~\cite{shiraishi2000designing,thorbeck2015formation}.

Low temperature operation of silicon quantum devices also presents difficulties in device simulation. Simulation programs based on solving the standard semiconductor equations (Poisson, drift-diffusion, continuity) suffer from poor convergence at low temperatures ($<$ 20\,K) owing to the increasing abruptness of functions describing the carrier populations. This limits the utility of industry standard TCAD simulation packages (such as Synopsys Sentaurus, Silvaco Victory, and similar). Methods of mitigating these convergence problems exist~\cite{mohiyaddin2019multiphysics}, however an alternative approach for advancing this capability is in the development of new code dedicated to semiconductor-based quantum device simulations. The emerging tools include both holistic solutions (e.g. QCAD~\cite{gao2012qcad}) and those integrated into pre-existing software (e.g. nemo5 in Silvaco TCAD, QuantumATK in Synopsys TCAD).

\subsubsection{The Transition from Noise Margins in Transistors to Decoherence in Spins}

The impact of noise in classical CMOS bits is only significant if it causes a fluctuation in electric fields and currents large enough to lead to a misidentification of the state of a transistor. In practice, CMOS devices have allowable operating margins that are much larger than the actual noise fluctuations, so that noise has no impact on the final result of a computation.

In qubits, noise has a much more serious impact, being associated with the decoherence time. Quantum error correction protocols impose an overhead on the number of qubits required that is set by the rate of errors, which is ultimately dictated by the noise in the device. This makes clear that noise minimization is a major goal for quantum computing and that the noise requirements are much more stringent for a quantum processor than for its classical counterpart. The issue of noise, its properties and microscopic origins are discussed further in Section 4.

\subsubsection{The Transition from Currents to Spins}

Classical transistors operate based on the movement of electrons (electric current), so that historically, significant engineering effort has targeted improved mobilities and threshold voltages. From the quantum information point of view, however, charge movement and electron position are not ideal variables due to their strong coupling to electrical noise~\cite{hayashi2003coherent}.

A common misconception is to assume that the fast decoherence of charge states caused by electric noise can be compensated by fast control of the qubit. In reality, error identification protocols create a demand to measure, classically process the result and create conditional control pulses on the qubits several times before decoherence dominates. This means that electric noise is more than just an engineering challenge -- it fundamentally hinders the use of charge states for quantum computing.

Shifting attention from charge to electronic spin states requires engineering the material stack with a focus on other fundamental properties. Examples include hyperfine coupling to $^{29}$Si nuclei and spin-orbit coupling. The relationship between these effects on spins and materials choices is discussed next.

\subsection{Active Silicon Layer}

\subsubsection{Geometry}

The first point of difference among the various CMOS foundry device platforms, as well as between foundries and academic laboratories, relates to the geometry of the active layer. Following decades of development focused on the performance of classical transistors, the industry has moved away from planar geometries in the past decade. However, the reasons behind this shift are based on the requirements of classical bits, and the ideal geometry for qubits is not yet clear.

Most electrostatic quantum dots made in university fabrication facilities are based on confining electrons against a planar interface and corralling them laterally with gate electrodes. The two most studied interfaces are the Si/SiO$_2$ interface and the Si/SiGe interface. We will later discuss in more detail the role of the barrier material.

Industrial platforms based on silicon FinFETs and fully-depleted silicon-on-insulator (FDSOI) nanowires have been used recently to isolate single electrons and control their spins~\cite{zwerver2021qubits,ciriano-tejel2021spin}. In these situations, the electron would be either accumulated at a corner of the nanowire or at the top surface of a FinFET, such that the electrostatic confinement can be achieved with a single gate wrapping around the transistor. This difference in confinement strategy in comparison to planar geometries has a large impact on qubit devices. Firstly, the quantum dots can be formed more efficiently with less gate electrodes required to create the confining electrostatic potential. On the other hand, the electrons are now in contact with the barrier interface (most often an amorphous oxide) in more than one direction, which has an impact on qubit performance and variability.

\begin{figure*}
  \includegraphics[width=\linewidth]{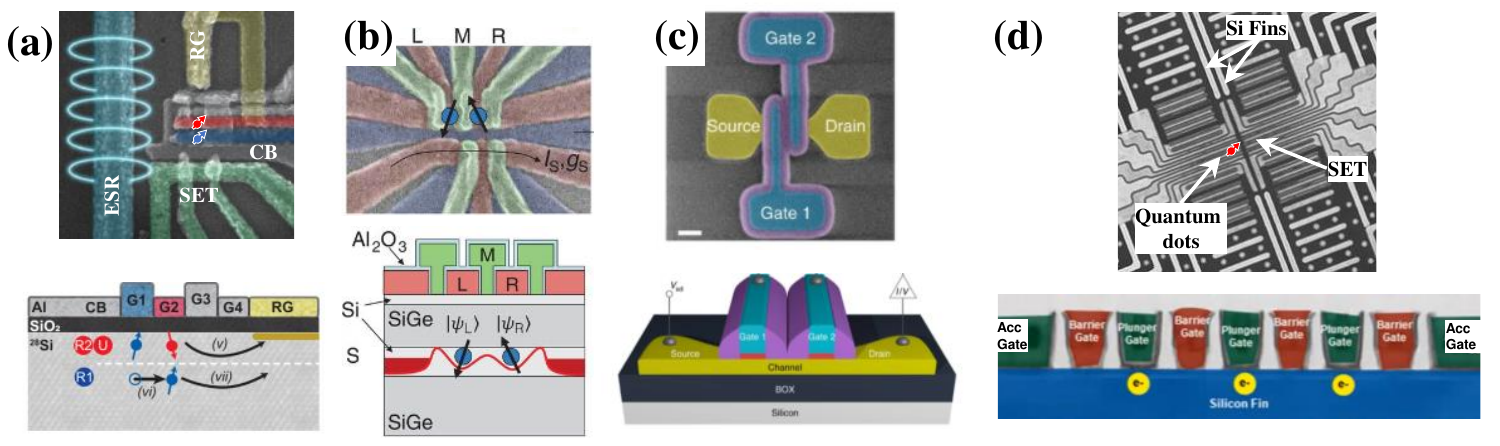}
  \caption{Comparison of four geometries and materials systems for silicon-based quantum dots. Scanning electron microscope (SEM) images of quantum dot devices and their respective schematic cross-sectional views: 
  (a) SiMOS planar device~\cite{huang2019fidelity}, 
  (b) Si/SiGe heterostructure device~\cite{zajac2018resonantly}, 
  (c) Silicon-on-insulator (SOI) nanowire~\cite{maurand2016cmos}, and 
  (d) SiMOS FinFET~\cite{zwerver2021qubits}.}
  \label{fig:boat1}
\end{figure*}

Figure~\ref{fig:boat1} compares examples of four different geometries. The first two devices in Figs.~\ref{fig:boat1}(a) and (b) were fabricated in academic research laboratories, and employ a planar active silicon layer with overlapping gate electrodes to provide vertical confinement of the electrons. In both cases, lateral confinement is achieved by gate electrodes that form electrostatic potential barriers controlling the tunnelling of electrons in and out of the dot. Devices shown in Figs.~\ref{fig:boat1}(c) and (d) were fabricated in industrial-scale foundries and in each case the electrons are confined by an oxide barrier both at the top and at least on one of its sides. These FDSOI nanowire and FinFET geometries for the active silicon layer allow for better performance in classical transistors, but their impact on quantum devices remains under investigation.

The dimensionality of the active silicon region plays a more substantial role than just the variation in confinement strategy. The arrangement of qubits in a two-dimensional lattice is a necessity for quantum error correction based on the surface code~\cite{fowler2009highthreshold}. The most direct strategy for this is positioning quantum dots in a two-dimensional array and tuning electrically the coupling between spins in neighbouring dots to create exchange coupling.

An exciting prospect for the fabrication of large scale quantum processors will be to access three-dimensional integration employing multiple layers of interconnects, and which could potentially allow for the stacking of qubits vertically~\cite{batude20153dvlsi}. Part of this can be achieved through vias and upper metal layers already employed in standard CMOS integration. This could create new pathways for error correcting codes (for instance by assembling data qubits and ancilla qubits in different layers). Alternatively, it can simply be used to improve the utilization of on-chip real estate and allow, for instance, for auxiliary devices such as amplifiers, resonators, reservoirs and single electron transistors to run in parallel with the qubit layer.

\subsubsection{Isotope Concentration}

The isotopic composition of the silicon substrate is an important factor for quantum processors which is not of relevance for classical transistors. Natural silicon has approximately a 4.7\,\% concentration of $^{29}$Si isotopes, which contain a non-zero nuclear spin, whereas two naturally abundant isotopes, $^{28}$Si and $^{30}$Si, are spinless. 

These nuclear spins are not stabilised by temperature because the Zeeman energy splitting is smaller than the thermal energy in dilution refrigerators. Moreover, these nuclear spins interact too weakly with the lattice to efficiently thermalise. This weak interaction can be leveraged to robustly encode quantum information in the nuclear spin states~\cite{hensen2020silicon}. However, a random distribution of these nuclei negatively impacts the electron spin coherence in a quantum dot.

The $^{29}$Si nuclei undergo \emph{flip-flop} interactions with each other, and these nuclear spin fluctuations couple through the hyperfine interaction with electron spins, causing shifts in the electron spin qubit resonance frequency~\cite{cywinski2009electron}. Even though this interaction can be quite weak, it is one of the leading sources of decoherence for spin qubits in silicon. We will further elaborate on the physical mechanism for this noise in Section 4.

Some effort in realising qubits in natural silicon has been undertaken~\cite{kawakami2014electrical, takeda2016faulttolerant}, but the coherence times are generally too short to contemplate fault-tolerant implementations. Better progress has been achieved with a purification level of 800\,ppm~\cite{yoneda2020coherent,yang2019silicon}. At this level, the number of $^{29}$Si nuclear spins in contact with the electron spin is reduced. This reduced number of spins does not reduce significantly the amplitude of the fluctuations in spin frequency~\cite{zhao2019singlespin}, but changes in this frequency become slow enough that they can be tracked and the qubit may be recalibrated dynamically~\cite{huang2019fidelity}. Frequency recalibration, however, imposes a significant time overhead and would be impractical for a large array of qubits.

Multiqubit systems will most likely require silicon purified to at most tens of parts per million of $^{29}$Si/$^{28}$Si, as already produced for the Avogadro project to redefine the kilogram metrology standard~\cite{brumfiel2010elemental,andreas2011determination}, which is and commercially available these days. This will reduce the chances of significant jumps in qubit frequency during operation, reducing the number of calibration steps required. 

\subsection{Barrier Insulators}

Insulators serve the double function of restricting the passage of currents from the gate electrodes to the active silicon channel and serving as a potential barrier to confine the electron wavefunction in the vertical direction. While these two purposes are related, the ideal material for each of these tasks might not be the same. 

Even with classical transistors, it is common to rely on a stack of different insulators to maximize the effectiveness of the fields generated by the gate, while maintaining the good chemical properties of the interface to minimize the density of interface traps. This is the case when a MOSFET has a thin silicon dioxide interface with the silicon channel and an additional high-$\kappa$ dielectric deposited prior to a metal gate.

We therefore differentiate between the two types of insulating materials and focus on their properties separately. We refer to them as a barrier insulator and a gate dielectric insulator. We highlight, however, that it is often possible to have the same material acting as both barrier and gate insulator.

\begin{figure*}
  \includegraphics[width=\linewidth]{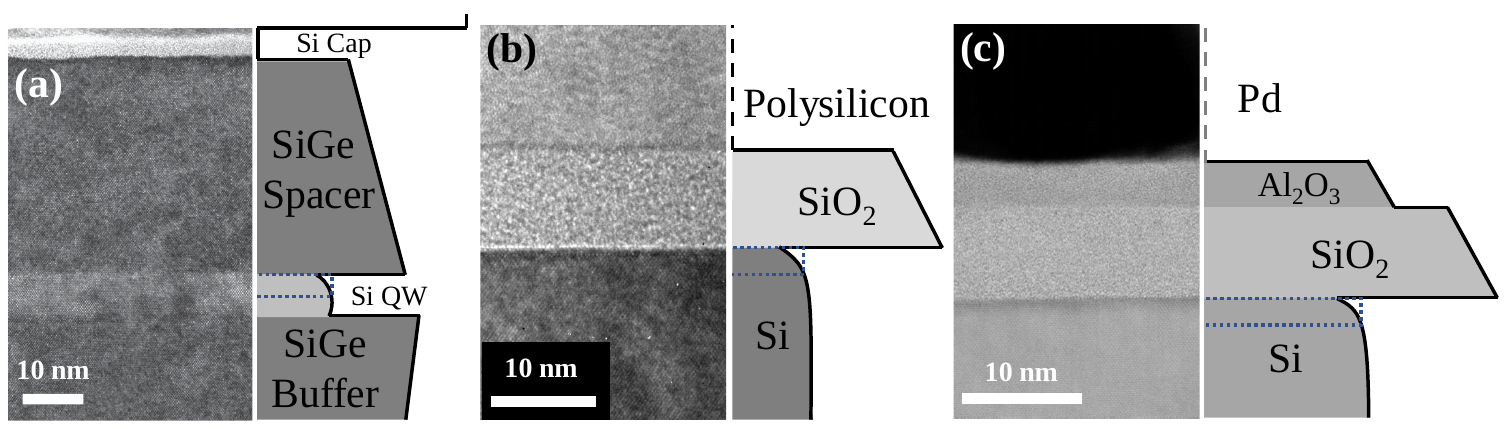}
  \caption{Materials stack for quantum dots. The TEM image of a cross section and the respective band structure diagram is shown for 
  (a) SiGe/Si/SiGe quantum well heterostructure~\cite{zajac2015reconfigurable}, 
  (b) SiMOS structure with polysilicon gate~\cite{carroll2015materials}, and 
  (c) SiMOS structure with Pd metal gate.}
  \label{fig:boat2}
\end{figure*}

Electrostatically defined quantum dots are formed by applying electric fields that create a potential minimum in the lateral directions (confinement potential) and a vertical electric field that compresses the electron against the interface between silicon and some other material that has an offset in its conduction band with regard to the conduction band of silicon.

What sets the choice of barrier insulator is firstly the thermodynamic stability with silicon (since it is in direct contact). After that, the most important characteristic for a barrier insulator is the quality of the formed interface. This can be quantified in terms of the interface roughness, homogeneity of the stoichiometry, surface density of defects and quantity of fluctuating charged defects within the barrier.

The most commonly adopted materials for barrier insulators used so far in silicon quantum dot literature are silicon dioxide and silicon-germanium alloys. Figure~\ref{fig:boat2} shows examples of material stacks adopting both options of barriers and how their conduction bands line up and bend to form confined electron states at the silicon/barrier interface.

\subsubsection{Silicon Dioxide as a Barrier Insulator}

The standard interface of choice for miniaturised industrial transistors is silicon with thermally-grown SiO$_2$ (Si/SiO$_2$). Despite the complex chemical nature of the oxide, this interface is the easiest to nanofabricate and presents one of the lowest densities of interface defects. For quantum computation, a few of its properties become very important. 

Firstly, the bulk SiO$_2$ is known to be amorphous. At the immediate vicinity of the Si/SiO$_2$ interface it is common to encounter small regions of trydimite, $\beta$-crystoballite and quartz~\cite{helms1994siliconsilicon}. Typically these materials combine to give a total band gap of the order of 6\,eV, with the conduction band being aligned 3\,eV higher than the conduction band of silicon (which represents the effective energy barrier for electron penetration into the oxide).

Transmission electron microscope images of the interface do not indicate that the interface has a recognizable crystal formation for the oxide, but there is some indirect evidence for a level of atomic ordering within the first few layers of oxide induced by the ordered silicon substrate. The most striking indication of this is the observation of piezoelectricity~\cite{lazovski2012detection}. A fully amorphized material should have no net measurable piezoelectricity, and this is indeed the case for bulk SiO$_2$. Nevertheless, for nanometric layers of oxide grown over silicon a detectable piezoelectric effect was reported. There are also some indications that piezoelectric phonons interact with spin qubits, causing the deterioration of its lifetime with temperature~\cite{yang2020operation}.

The coefficients of thermal expansion of silicon and its oxide differ significantly, with silicon contracting five times faster than silica at room temperature. At cryogenic temperatures, this can induce significant uniaxial tensile strain on the silicon layer. Further investigation of strain fields in cryogenic samples are needed to elucidate the extent of this effect.

The thickness of the oxide layer plays various roles. At just a few nanometers, its role as a barrier to avoid tunnelling from electrodes to the active channel is already accomplished for typical electric fields adopted in quantum dot operation. From the point of view of quantum dot properties, the thickness of the oxide also sets its lateral size. An oxide that is thinner than the gate width will lead to a dot formation in which the lateral confinement length is set by the gate lateral size (except in the presence of disorder). If the barrier material used is much thicker, the distance from gate to silicon channel creates a smoother, wider confinement potential.

Both the large electric field induced in the channel and the smaller dot sizes result in a larger valley splitting for SiO$_2$ barriers than what is seen, for instance, for SiGe alloy barriers. This is because valley splitting is negatively impacted by interface roughness due to valley interference effects~\cite{ibberson2018electricfield}. The roughness of the Si/SiO$_2$ interface has a larger amplitude over longer lateral distance scales, in a self-similar fashion. Smaller dots create electron wavefunctions that do not spread too far, resulting in an effectively sharper interface and more efficient valley splitting.

The roughness of the interface profile also has a direct impact on the electron spin. While the bulk silicon lattice has inversion symmetry, resulting in zero spin-orbit effect with Dresselhaus symmetry, an atomically sharp interface breaks this symmetry and creates a preferential in-plane direction for the spin-orbit field~\cite{ruskov2018electron}. This is an important effect because it is dependent on the atomic scale structure of the interface, and as such it is directly impacted by roughness variability between different dots.

A thin oxide also has advantages from the quantum control point of view, due to the large gate-dot capacitance. With a larger capacitive coupling between gate electrodes and the dot, gate-based dispersive readout~\cite{betz2015dispersively} and spin-photon coupling~\cite{mi2018coherent,samkharadze2018strong} become more efficient. In general the electric control of spins, such as in electrically-driven spin resonance, is positively impacted by a thinner oxide, owing to a larger dot-gate coupling~\cite{golovach2006electricdipoleinduced} (referred to as the lever arm).

The quality of the oxide is a key parameter for classical transistors and is even more important for qubits. Fluctuating charged defects create random telegraph noise that leads to spin dephasing. Moreover, the telegraph noise is picked up by the electrometer (SET or resonator), contaminating the measurement with noise and requiring longer integration times to attain sufficient signal-to-noise ratios. This is a very important limitation since it reduces the measurement bandwidth -- a bottleneck for the repetitive measurements required for quantum error correction.

We briefly mention that some SOI technologies include the use of buried oxides, which could potentially lead to significant differences in qubit operation. In particular, there are some results that indicate that a buried oxide could lead to extremely large valley-orbit coupling~\cite{takashina2006valley}. This could potentially be explained in terms of the formation of band tail interface states (Tamm/Shockley states), which depend on the details of the interface chemistry~\cite{saraiva2010extended}. No qubits have ever been measured on a quantum dot that uses these types of interfaces, however.

\subsubsection{Epitaxial \texorpdfstring{Si$_{1-x}$Ge$_x$}{SiGe} as a Barrier Insulator}

Another material used in industry is Si$_{1-x}$Ge$_x$. This material has been used for the active channel for certain classes of transistor devices, but it can also be used as a barrier insulator for the accumulation of electrons at a Si/Si$_{1-x}$Ge$_x$ interface. Devices are typically fabricated by epitaxially growing strained Si on a substrate of relaxed Si$_{1-x}$Ge$_x$, such that the conduction band of silicon is shifted to lower energies due to the resulting uniaxial tensile strain~\cite{schaffler1997highmobility}.

Because both Si and Ge atoms have the same valence configuration, the electronic properties of the alloy are qualitatively similar to either bulk silicon ($x$ small) or bulk germanium ($x$ close to 1). This means that the relaxed alloy will have a conduction band that is offset with regard to the strained silicon quantum well. This conduction band offset can be as large as hundreds of meV.

The stress created by typical concentrations of germanium ($x\approx0.3$) in silicon can lead to the formation of topological defects, and only thin quantum wells are stable enough to be fabricated in this manner. This is in contrast to Si/SiO$_2$ interfaces, which can be used for an active silicon layer of arbitrary thickness.

\begin{figure*}
\centering
  \includegraphics[width=0.7\linewidth]{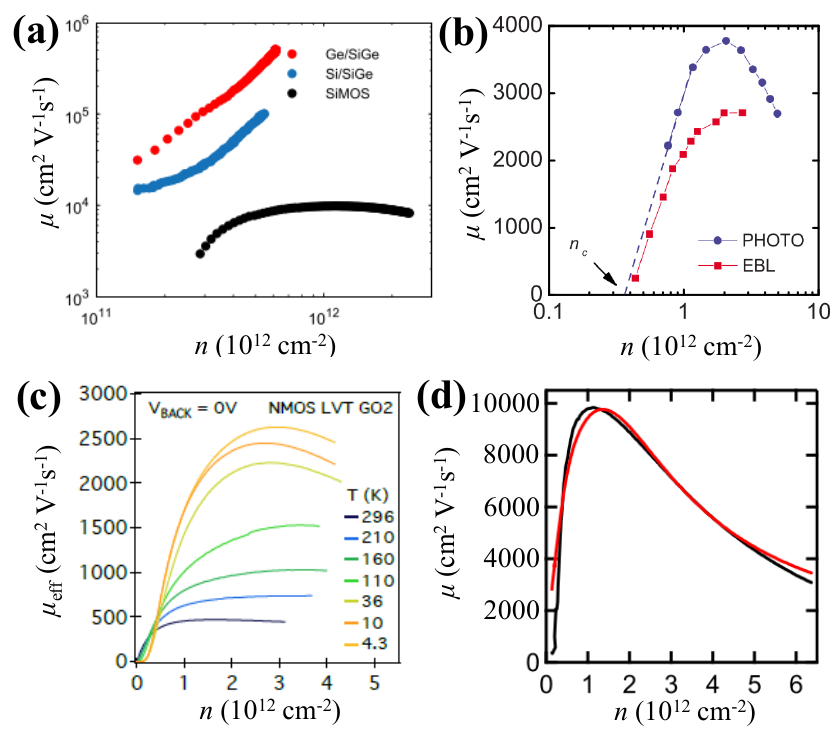}
  \caption{Mobility measurements as a function of carrier density in different silicon platforms. 
  (a) A comparison between devices based on SiMOS and Si/SiGe wafers (as well as Ge/SiGe), with the SiGe heterostructures exhibiting higher maximum mobility and lower critical density~\cite{lawrie2020quantum}. 
  (b) SiMOS Hall bar device fabricated in an academic cleanroom via photolithography gives a higher peak mobility and lower critical density compared to a device fabricated using electron beam lithography~\cite{WillemsvanBeveren2010overlappinggate}. 
  (c) Effective mobilities of NMOS as a function of temperature ranging from room temperature to 4.3\,K~\cite{bohuslavskyi2018cryogenic}. 
  (d) Hall-bar transistor made on industrial $^{28}$Si/$^{28}$SiO$_2$ measured at 1.7\,K with a peak mobility of 9800\,cm$^2$/Vs and a critical density, $n_\mathrm{c}$ of 1.75 $\times$ 10$^{11}$\, cm$^{-2}$~\cite{Sabbagh2019quantum}.}
  \label{fig:boat3}
\end{figure*}

Among the advantages of SiGe over SiO$_2$ barriers are the higher mobilities for a two-dimensional electron gas confined at the barrier interface and lower percolation densities~\cite{lawrie2020quantum}. Figure~\ref{fig:boat3}(a) shows a comparison of the mobilities of electrons confined at the interfaces of Si/SiGe and Si/SiO$_2$ devices. The higher mobilities achieved using Si/SiGe interfaces result from the good lattice match between Si and Si$_{1-x}$Ge$_x$ for a stoichiometric fraction of up to $x=0.3$, meaning that the number of defects at the interface is small in comparison with the Si/SiO$_2$ interface. Thermally grown SiO$_2$ is amorphous over large length scales and, therefore, not lattice matched. SiGe also has a lower density of fixed charge defects in comparison with SiO$_2$, leading to a lower number of two-level fluctuators, and therefore reduced random telegraph noise.

The larger distance between the gate electrodes and the active silicon layer for SiGe barriers as compared with SiO$_2$ barriers also impacts the size of the quantum dot, leading to a larger spread of the electronic wavefunction laterally. While this is an advantage for relaxing the fabrication demands for the gate electrode layout, it also exposes the qubit to undesirable effects. Examples include the reduction of the capacitive coupling between gates and dots (which limits the effectiveness of gate-based readout and electrically-driven resonance) and a reduction in the average valley splitting. Typical valley splitting energies in Si/SiGe quantum dots are in the range 0-0.4 meV, whereas in Si/SiO$_2$ dots the range is typically  0.2-1.0 meV. Si/SiGe structures are also restricted to planar device geometries, limiting the use of the full range of complex processes that are at disposal for MOS devices.

\subsection{Gate Materials, Gate Oxides and Dielectric Insulators}

The demand for a dense arrangement of electrodes in order to form highly controllable potential profiles for quantum dot devices pushes the technology of gate fabrication to its limits in quantum dot devices. One of the most successful approaches employs a multilayer gate stack with very thin oxides between gates~\cite{lim2009observation, yang2014charge}.

The material used for insulating gates from each other and avoiding leakage from the gates to the silicon layer can be different from the material used as a barrier in direct contact with silicon. In industrial applications, silicon dioxide was historically adopted for both purposes, especially in the context of gates made from polysilicon.

Metallically-doped polysilicon was largely adopted as the industry standard for MOSFET gate electrodes for decades due to the resulting low threshold voltages (they have essentially the same work function as n-Si), which relaxes the constraints on the gate oxide (operating at lower fields, leakage is more easily avoided). This allowed the use of a single type of oxide (SiO$_2$) to serve as both gate oxide and barrier for the electrons.

Two issues with polysilicon have steered the industry away from it for highly-scaled CMOS transistor nodes. Firstly, operation of polysilicon-based transistors at ultra-high frequencies is limited due to its resistivity (even at very high doping)~\cite{shenai1990gateresistancelimited}. Perhaps more importantly, polysilicon suffers from a depletion effect. This is a reference to the fact that even at very high doping, a finite depletion layer forms at the outer surface of the gate~\cite{lo1999modeling}. This causes two problems -- the effective capacitance from gate to the active silicon layer is reduced; and the very small volume of the depletion region is subject to the natural variability in doping density, leading to a statistical dispersion in device characteristics such as threshold voltage.

These difficulties have led to the development of gate stacks using high-$\kappa$ dielectric insulators and metal gates (commonly referred to as HKMG technology~\cite{robertson2015highk}). Examples of metals widely used (especially starting after the 45/32\,nm nodes) include Ta, TaN, Nb, Al, W and Pd, while the most common high-k dielectric in use are HfO$_2$, PolySiON and others.

These high-$\kappa$ dielectric insulators can be made thicker without sacrificing the capacitive coupling between the gate and the electrons in the active layer, minimizing channel-to-drain leakage. With a larger repertoire of dielectric materials, the possibilities in terms of process engineering and resulting densities of defects have greatly advanced in the past decades. One example is the possibility to use noble metals, which do not spontaneously form native oxides but can be capped off by oxides fabricated by atomic layer deposition. Another example of the flexibility of HKMG processes is the possibility to create overlapping gate layouts.

We note that specific choices of HKMG materials are seldom revealed by foundries, being a key element of industrial and intellectual property for chip makers. In the context of academic research, the most widely explored materials for quantum dot fabrication have been so far Al and Pd. 

Aluminium, being the material first used in the CMOS industry, has had the head start. It has excellent compatibility to most MOS processes, forms a good quality thermal oxide and has good conductivity for high frequency applications. On the other hand, depending on process parameters, aluminium can form large grains, which limits its usage for gates with very small features. Moreover, aluminium has a significantly different coefficient of thermal expansion when compared to silicon, which creates an inhomogeneous strain profile under the gate in cryogenic conditions~\cite{thorbeck2015formation}.

Palladium gate electrodes can be made much smaller without running into the issue of grain sizes being comparable to gate widths~\cite{brauns2018palladium}. It also has a coefficient of thermal expansion that is not so strongly mismatched to that of silicon, reducing the inhomogeneity of the strain induced by the gate stack. The main shortcomings of Pd are the relatively low conductivity and its lack of a natural oxide. This means that Pd gates must be insulated using some deposited oxide.

An important issue with nanoscale metal gates is the lithography technique adopted. In academic research laboratories, the only affordable method to fabricate feature sizes down to 20\,nm is electron beam lithography. This is an inherently slow process with low throughput because the electron beam exposure is point-by-point. In foundries, deep ultraviolet (DUV) or extreme ultraviolet (EUV) lithography are used to pattern gates in an entire device or multiple devices in a single exposure. Another disadvantage of electron beams is the exposure to high-energy impacts that could damage the heterostructure. Figure~\ref{fig:boat3}(b) shows an example of the impact of this damage on the transport performance of a Hall bar.

Figures~\ref{fig:boat3}(c) and (d) show the impact of temperature and isotopic purification on the mobilities, respectively. Temperature effects are well studied in the context of classical transistor since they limit the mobilities at high carrier densities $n$. It is hard, however, to draw any conclusion on the impact of this degradation of mobility on localised electron in quantum dots. Isotopic purification, on the other hand, does not influence mobility significantly, but has important implications for spin qubits. These are good examples of the difficulty in converting traditional metrics of materials performance to figures of merit for qubit operation.

\begin{figure*}
  \includegraphics[width=\linewidth]{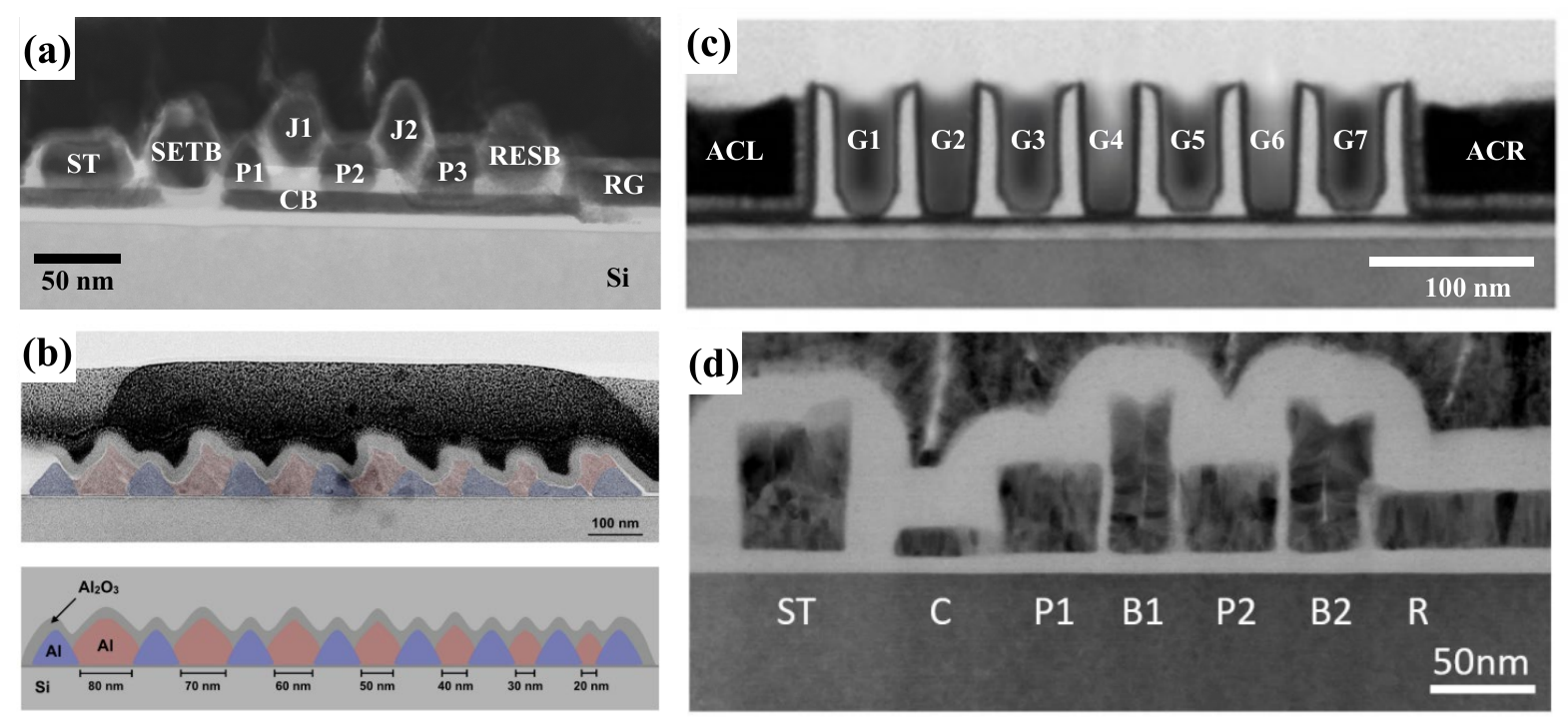}
  \caption{Transmission electron microscope (TEM) images of gate stack in quantum dot devices made in academic laboratories 
  (a) UNSW, and 
  (b) University of Wisconsin~\cite{dodson2020fabrication}; and in foundries 
  (c) Intel~\cite{zwerver2021qubits}, 
  (d) IMEC~\cite{li2020flexible}.}
  \label{fig:boat4}
\end{figure*}

The uniformity and geometrical conformity of gates produced in advanced VLSI foundries is also an important advantage that results from the decades of investment in classical transistor miniaturization. Figure~\ref{fig:boat4} shows a comparison between cross sections of devices from academic and industrial facilities, for example.

\subsection{Auxiliary Materials}

In its simplest incarnation, a spin-based quantum processor chip can be completely fabricated using only traditional CMOS materials and processes. On the other hand, the integration of quantum materials may provide added capabilities of long-range coupling between qubits, fast readout, spin-photon conversion and fast electric spin manipulation. These applications are facilitated by materials with special properties, such as superconductivity, magnetism, and others.

\subsubsection{Superconductors}

Superconducting materials are the natural choice for the creation of electromagnetic microwave cavities with high quality factors ($Q$ factors). These high $Q$ factors are necessary for multiple applications. We briefly discuss some of them next.

The first potential application of superconducting cavities is their use as resonators for dispersive qubit readout~\cite{west2019gatebased}. Due to the very small capacitance associated with single electron transitions, it is important to use a resonator with very narrow resonances, such as a high-$Q$ superconducting cavity~\cite{wallraff2005approaching}.

Another application of these cavities is to harbour long-lived photons, which can be used for spin-spin coupling mediated by a photon~\cite{mi2018coherent,samkharadze2018strong}. Note that photons interact very weakly with spins, typically through engineered inhomogeneous magnetic fields. Therefore, it is important to guarantee that the spin-photon interaction lasts long enough for the spin to transfer its quantum state to the electromagnetic modes of the cavity.

The conditions imposed by fabrication and operations of spin qubits in silicon restrict the choices of superconducting materials significantly. Firstly, for qubits based on a single electron spin, it is common to adopt magnetic fields in the range of 0.2\,T to 2\,T. Most traditional superconducting materials have critical fields well below these values, including aluminium which is the favourite superconducting material for transmon-based qubits~\cite{lisenfeld2019electric}. Restricting the magnetic field to be in-plane and adopting superconducting materials deposited as thin films (with thickness small compared to the coherence length of the superconductor), a larger magnetic field can be tolerated~\cite{litovchenko1971critical}. Type-II superconductors, such as Nb, NbN and NbTiN, can also be used in order to extend the range of fields that can be adopted. Research on the impact of vortices on the spin-photon system is currently immature, however.

Another critical issue is the temperature of operation. As mentioned before, a key advantage of qubits based on spins in silicon is their robustness against temperature. The advantages resulting from this relaxed temperature of operation come at the cost of limiting the range of superconductors to those that resist temperatures up to a few kelvin.

From the fabrication point of view, superconducting materials must be compliant with CMOS fabrication rules. The most restrictive fabrication step is the high temperature annealing in a forming gas, typically at around 450$^\circ$ C, which is a key step for repairing the silicon surface after all the chemical processes have been finalised. For example, Nb and its alloys are not compatible with a high temperature annealing step. This means that an on-chip resonator based on this material has to be processed after the rest of the material stack is fabricated and annealed.

\subsubsection{Magnetic Materials}

A common strategy to control spin qubits in quantum dots is to impose an oscillatory movement of the quantum dot within an inhomogeneous d.c. magnetic field, creating an effectively time-dependent magnetic field in the reference system of the spin. In order to create a synthetic inhomogeneous magnetic field, it is necessary to deposit a micromagnet near the dot region~\cite{pioro-ladriere2008electrically}.

Some of the challenges imposed by the presence of a magnet include the scalability of this technology for a dense array of quantum dots, the exposure of the qubit to electric noise and potentially the magnetic noise introduced by the demagnetisation of the micromagnet with larger temperatures and lower external magnetic fields.

From the fabrication point of view, there is no fundamental difficulty in depositing a magnetic material. But from the point of view of the logistics of a factory operation, magnetic materials are considered to be highly contaminating and are only incorporated in later stages of the device processing to avoid introducing undesired magnetic impurities in the production line.

It is worth mentioning that a superconducting film can also create a magnetic field gradient resulting from the Meissner effect (superconducting diamagnetism)~\cite{underwood2017coherent}. This could potentially be leveraged to create an effect similar to the field of a micromagnet without needing to introduce a magnetic material in the process. Whether there will be an advantage from the point of view of process engineering and if a material suitable for the operation conditions can be found remains to be investigated.

\subsubsection{Other Advanced Materials}

The extremely large range of materials at the disposal of advanced fabrication facilities is only now starting to be explored in integration with qubit devices.

One example is the use of quantum paraelectric materials with ultra-high permittivity for the fabrication of dielectric cavities~\cite{vahapoglu2021singleelectron}, which can be used for creating global microwave magnetic fields without introducing significant electric noise.

An alternative for superconducting cavities are nanoacoustic resonators that are fabricated with suspended nanobeams~\cite{mohammadi2009highq}. These devices create phonon modes with extremely long lifetimes. With $Q$ factors many orders of magnitude superior to electromagnetic cavities, they can potentially create a pathway for long-range coupling of spin qubits if coupled to quantum dots.

Another pathway for engineering acoustic waves are the piezoelectric materials. They can be used to induce surface acoustic waves and coherently transport electrons between remote quantum dots. This technology has been demonstrated in GaAs-based electrostatic quantum dots~\cite{hermelin2011electrons} and could be extended for use in silicon by either leveraging the residual piezoelectricity of the Si/SiO$_2$ interface or by depositing a layer of piezoelectric material atop a thin barrier oxide.

It is not hard to imagine potential applications for many types of quantum materials. The difficulty is that qubits are very fragile and it is important to confirm whether these materials introduce new forms of noise or if the fabrication process associated with these materials degrade the properties of the quantum dots in a significant way. Moreover, the quantum material needs to be compatible with the strict fabrication process rules for CMOS device manufacture..

\section{Noise Sources and Quantum Errors}

Noise impacts quantum processors in a more dramatic way than it does to their classical counterparts. Understanding its microscopic origins can help in devising new quantum-grade materials and processes.

\subsection{Characteristics of Noise and its Impact on Quantum Error Correction}

From the point of view of quantum error correction, most forms of noise are interpreted as a random transformation that may occur to a qubit with a certain probability, regardless of its microscopic origin. The most typical example is a qubit that can be impacted by a stray electromagnetic field, causing it to either precess faster or slower with respect to its neighbours, for instance. This is referred to as a phase-flip error~\cite{preskill1998reliable}.

If this type of error was the only one acting on a qubit, error correction would be very simple, accomplished by merely encoding the qubit in the quantum state of a set of qubits (which we call a logical qubit) and determining the state of the logical qubit by a majority vote. The situation is complicated by the fact that other errors, such as the bit-flip error (random rotation about the $x$-axis, for instance) may also occur. The X and Z components of a spin obey an uncertainty principle, limiting our ability to use a repetition code~\cite{wootton2018repetition} and majority vote strategy to distinguish these two types of errors, since the measurement of one of these variables has a back-action on the other.

More generally, the quantum information can be protected by encoding it in a delocalised collective state of physical qubits. The rules for converting these collective states into quantum information, as well as for detecting and stabilizing its information, are what we call a code. We refer to a code as fully quantum when it is able to identify and correct errors in non-commuting observables (X and Z in this example). The most widely studied version of these codes is the surface code~\cite{fowler2012surface}. Among its advantages, it relies on an easily imaginable two-dimensional array of qubits, with interactions limited to nearest neighbours only. It uses the two-dimensional topology of the qubit arrangement to encode information in the form of two interlaced square sublattices, each responsible for flagging one type of error. Repetitive measurements in a subset of the qubits in the array (called ancilla qubits) allow the triangulation of a potential qubit error. Conditional on these measurements, a correction pulse is applied at the position where the error occurred.

This code also deals with measurement errors. One can think of the state of the two-dimensional checkerboard pattern formed by a surface code in a given cycle of the error correcting scheme as a single slice in a three-dimensional stack of surfaces. The measurement error is then identified by triangulating it not in space, but in time.

In reality, one may generalise this encoding scheme to deal with more complex types of errors, such as SWAP errors (when qubits swap places) and even more dramatic qubit-loss errors. The particular operation that the noise source creates on the qubit is secondary. The main challenge refers to how these errors manifest.

\subsection{Requirements for a Noise Source to be Correctable}

The successive entangling gates necessary to create the collective states that are measured by the ancillas create a natural propagation of the errors in the qubit array. Since the error correction scheme relies on triangulating the qubit that caused the error, it is important that every instance of an error is well isolated from other errors that might have happened in the same cycle. If two errors occur in close proximity, the identification of the qubits that caused the original error is compromised.

\begin{figure*}
  \includegraphics[width=\linewidth]{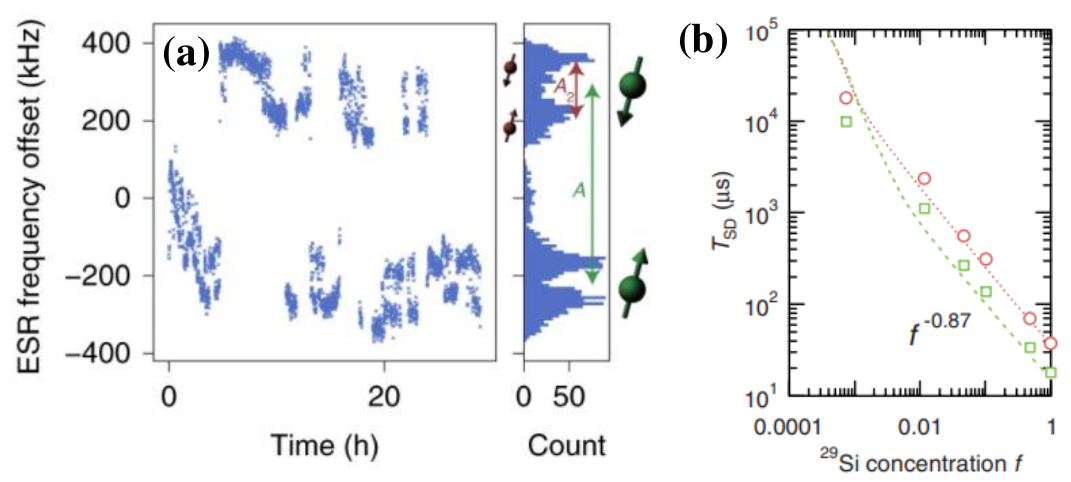}
  \caption{Impact of magnetic noise on spin qubits. 
  (a) Monitoring of spin resonance frequency of a single-electron SiMOS quantum dot qubit over a timescale of 24 hours. The histogram at right shows the presence of two $^{29}$Si nuclei with distinguishable hyperfine coupling~\cite{hensen2020silicon}. 
  (b) Nuclear-induced spectral diffusion time T$_\mathrm{SD}$ as a function of $^{29}$Si concentration, $f$~\cite{abe2010electron}.}
  \label{fig:mag-noise}
\end{figure*}

The first demand that this sets is that the probability for any random qubit to have an error occurring is low enough, such that these pair occurrences are rare~\cite{steane2003overhead}. This sets a threshold for qubit gate errors, as well as measurement and initialisation. 

The second, most striking demand is that errors are not strongly correlated spatiotemporally~\cite{clemens2004quantum}. Examples of spatially correlated noise would be a stray field that acts on two qubits simultaneously, causing the same error on both~\cite{boter2020spatial}. Temporal correlation could, for example, represent a field that acts on a qubit at a certain cycle, but does not immediately vanish and acts again on the qubit in a subsequent cycle~\cite{dehollain2016optimization}.

It is therefore important to characterise these noise correlations and develop methods to mitigate them. Luckily, correlated noise can be circumvented by echo techniques, as well as spectator qubit strategies.

\subsection{Magnetic Noise}

Magnetic noise refers to stray, uncontrolled magnetic fields created either by magnetism in the materials or by the cryogenic setup itself. These fields directly couple to the qubit and almost exclusively lead to phase-flip errors. That is because typically the time variation in a magnetic field is quite slow. That means that the stray magnetic field only sums to the external magnetic field, redefining the qubit precession frequency and potentially its quantisation axis. It also means that this is a strongly time-correlated noise.

Examples of prominent sources of slow magnetic noise include:

\begin{itemize}
    \item The contact hyperfine coupling with the magnetic moment of nuclear spins, in particular the naturally abundant $^{29}$Si~\cite{cywinski2009electron}, but also potentially from atoms of oxygen and from metal gates, in the form of long-range dipolar coupling;
    \item Vibration effects on the magnet used to create the quantisation field~\cite{kalra2016vibrationinduced};
    \item Variations in the diamagnetism of superconducting materials in the gate stack (Meissner effect) when operating near the superconducting transition point;
    \item Thermal fluctuations of the magnetization of an on-chip magnet~\cite{neumann2015simulation};
    \item Magnetic moments of Pb centres and other paramagnetic defects in the material stack~\cite{desousa2007danglingbond}.
\end{itemize}
    
Some fast magnetic noise can also occur. This could be the result of imperfections in the microwave source used to drive the qubits, its mode of delivery, etc.

Note that the most important source of magnetic noise that greatly impacts spin qubits is the hyperfine coupling to nuclear spins. This noise source can be significantly improved by isotopic purification, after which the residual source of errors becomes the electrical noise.
More importantly, the nuclear spins are very stable, such that echo techniques are very efficient at correcting this type of error. It is also possible to feedback the qubit frequency shifts in real time.

\begin{figure*}
  \includegraphics[width=\linewidth]{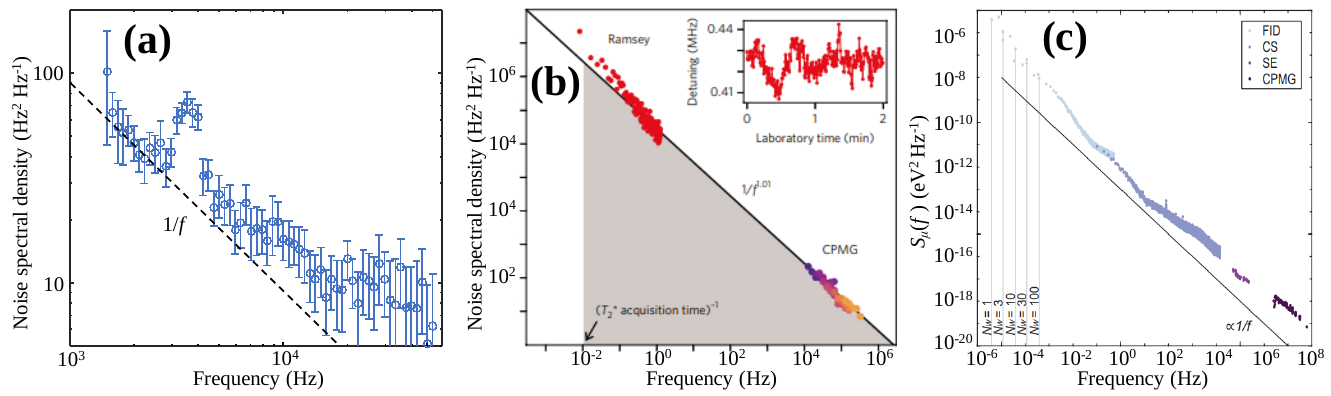}
  \caption{Impact of electric noise on qubits, as measured in a 
  (a) SiMOS quantum dot without a micromagnet~\cite{chan2018assessment}, 
  (b) SiGe quantum dot with a micromagnet, and~\cite{yoneda2018quantumdot} 
  (c) SiGe singlet-triplet qubit (converted to electrostatic potential noise)~\cite{connors2021chargenoise}. }
  \label{fig:e-noise}
\end{figure*}

Figure~\ref{fig:mag-noise}(a) shows the result of tracking a spin qubit in a SiMOS quantum dot device over a timescale of hours. Two distinctive nuclear spins are identifiable in this example, since at 800\,ppm concentration of $^{29}$Si/$^{28}$Si only a handful of nuclear spins are strongly coupled to the electron spin in a small quantum dot. The overall improvement in qubit coherence due to isotopic purification is shown for an ensemble of electron spins in Fig.~\ref{fig:mag-noise}(b). Compared to typical gate operation times (which range from 0.1 to a few microseconds), the spin dephasing time $T_{SD}$ is only long enough to allow a large number of operations if the concentration of $^{29}$Si is reduced.

Moreover, hyperfine coupling is a contact interaction, which means that this source of noise is not spatially correlated. The long-range dipolar coupling is, however, strongly delocalised and can create noise correlations between errors in different qubits.

\subsection{Electrical Noise}

Electrical noise is a significant issue in all semiconductor devices, especially in the presence of dielectric materials. While spins are not directly affected by the noise in electric fields, their spin-orbit effect creates a small yet relevant coupling. If the d.c. magnetic field is homogeneous, the most relevant way in which electric fields impact spin qubits in silicon quantum dots is through the renormalisation of the electron g factor in the presence of an interface~\cite{veldhorst2015spinorbit}. In the presence of an inhomogeneous magnetic field due to a magnet, for instance, the lateral displacement of the dot induced by electric noise also creates an effective fluctuation in the magnetic field.

The electric noise will change the intensity of the field pressing the electron against the oxide interface. Moreover, it will move the electron laterally, which changes the spin-orbit effect experienced by the electron due to roughness-induced variations in the interface atomic profile.

Sources of electric noise include:

\begin{itemize}
    \item The electrical setup, which introduces voltage fluctuations at the gate electrodes;
    \item The trapping and untrapping of electrons from the silicon layer to the barrier oxide interface trap states that form due to imperfections in the oxidation process~\cite{cheng1977electronic};
    \item The movement of atoms in the oxides between two configurations that are equally energetically favourable~\cite{helms1994siliconsilicon}.
\end{itemize}

The first form of noise is typically associated with the control setup at elevated temperatures, and can be improved through the use of CMOS cryo-controllers~\cite{hornibrook2015cryogenic,patra2018cryocmos}. Moreover, a compromise between filtering and speed of operations can improve this type of noise.

All other forms of noise are the result of the choice of material stack and process steps. They all have the same phenomenology -- they are two-level systems that have individual natural transition frequencies, and their fields have a long range. This means that the qubits are typically subject to the summed effect of many two level systems, which add up to a total noise profile that has a power spectral density following a 1/$f$ law~\cite{burkard2009nonmarkovian}. Direct measurements of this effect in qubit devices in silicon are shown in Fig.~\ref{fig:e-noise}. Note that the conversion between oxide electric noise and spin noise depends on details of the device layout, such as the presence of micromagnets, the distance between the active layer and the oxide, and the direction of the magnetic field with regard to the crystal structure.

This effect is extremely well studied in the silicon microelectronics device community. In the context of qubits, it leads to some particular features. For instance, the nanoscopic nature of the qubits often means that one of the two-level systems is in closer proximity than others, adding certain characteristic noise frequencies that can be easily distinguished in a noise spectroscopy experiment. 

\section{Qubit Variability and Large Scale Testing}

Mass fabrication of qubit devices will require the development for fast automated characterisation of quantum dots and qubits~\cite{pillarisetty2019high}. This is a significant shift from the current status of the field, in which the tuning of quantum dots and detection of spin transitions is mostly done by hand.

Moreover, testing the impact of changes in the device fabrication process on qubit performance is currently a lengthy process. The correlation between classical semiconductor devices metrics (such as mobility) and spin qubit performances is a current topic of investigation.

This means that screening for good potential qubit devices requires tests performed at milikelvin temperatures, which typically consists of cooling down devices in a dilution refrigerator, limiting test rates to one device every few days, unless cryo-multiplexers or cryo-probestations~\cite{pillarisetty2019high,paqueletwuetz2020multiplexed} are used. As the community ramps up the knowledge of the correlation between device characteristics at higher temperatures (liquid helium or even room temperature) and qubit performance, the risks associated to changes in fabrication process will be mitigated.

\section{Conclusion}

Transitioning from a history of university-led research to a future leveraging the development capabilities of advanced VLSI foundries, spin qubit research and development is experiencing a rapid increase in interest and commercial investment~\cite{gonzalez-zalba2020scaling}. While the CMOS industry has built up a formidable store of knowledge over the past decades regarding material properties and their impact on electronic devices, classical transistor metrics do not entirely determine the success of a semiconductor process in creating good spin qubits. New device parameters become relevant at the scale of single electrons in cryogenic conditions. 

The role of scientific advances in this field will also progress from basic elementary demonstrations to more advanced topics, such as large scale integration from a systems perspective, quantum error correction and auxiliary technologies such as quantum links and quantum memories. While there is strong evidence that silicon may serve as an ideal material for the quantum information revolution, as it has for its classical counterpart, there remains important materials science research to be done to enable large scale quantum processing to reach maturity.

\section*{Acknowledgements} 
We acknowledge support from the Australian Research Council (FL190100167, CE170100012 and LE160100069), the US Army Research Office (W911NF-17-1-0198), the NSW Node of the Australian National Fabrication Facility and UNSW Sydney. The views and conclusions contained in this document are those of the authors and should not be interpreted as representing the official policies, either expressed or implied, of the Army Research Office or the US Government. The US Government is authorized to reproduce and distribute reprints for Government purposes notwithstanding any copyright notation herein.
\medskip

\bibliographystyle{MSP}
\bibliography{refs}

\end{document}